\documentclass{article}

\usepackage{microtype}
\usepackage{graphicx}
\usepackage{subfigure}
\usepackage{booktabs} 

\usepackage{hyperref}

\usepackage[accepted]{icml2018}

\icmltitlerunning{Artificial Intelligence Techniques for Security Vulnerability Prevention}

\usepackage{amsmath,amsfonts,amsbsy,amssymb}
\usepackage{balance}
\usepackage{paralist}
\usepackage{enumitem}
\usepackage{listings}
\usepackage{setspace}
\usepackage{multirow}
\usepackage{xspace}
\usepackage{array}

\usepackage{fancyhdr}

\newcolumntype{H}{>{\setbox0=\hbox\bgroup}c<{\egroup}@{}}
\setlist{nolistsep}
\setlist[itemize,1]{leftmargin=\dimexpr 26pt-.25in}

\begin{document}

\twocolumn[
\icmltitle{Artificial Intelligence Techniques for Security Vulnerability Prevention}

\icmlsetsymbol{equal}{*}


\begin{center}
Steve Kommrusch \\
steve.kommrusch@gmail.com
\end{center}
\begin{center}
Colorado State University \\
Department of Computer Science \\
Fort Collins, CO, 80525 USA \\
April 2018
\end{center}


\icmlcorrespondingauthor{}{}


\vskip 0.3in
]



\printAffiliationsAndNotice{}  

\begin{abstract}

Computer security has been a concern for decades and artificial intelligence techniques
have been applied to the area for nearly as long. Most of the techniques are being applied
to the detection of attacks to running systems, but recent improvements in machine learning
(for example, in natural language processing) have enabled the opportunity to process
software and specifications to detect vulnerabilities in a system before it is deployed. 
This paper presents a survey of artificial intelligence techniques (including machine 
learning) to detect or repair security vulnerabilities before product introduction.  
In the surveyed papers, techniques are presented for using NLP to analyze requirements
documents for security standard completeness, performing neural fuzz testing of software,
generating exploits to detect risk, and more.  We categorize current techniques into 3 
groups: vulnerability detection, vulnerability repair, and specification analysis.
Generally, while AI techniques have become quite useful in this area, we show that AI
techniques still tend to be limited in scope, providing a collection of tools which can
augment but not replace careful system development to reduce vulnerability risks.  

\textbf{Keywords}: artificial intelligence, machine learning, security, vulnerabilities, software

\end{abstract}

\section{Introduction}
\label{sec:introduction}

Machine learning has been expanding in scope in recent years to the point where
meaningful natural language processing (NLP), automated testing, and code analysis
can be utilized. Additionally, more tradition artificial intelligence (AI) techniques
such as support vector machines, genetic algorithms, and inference engines can
be applied to code before it is used in a setting where it may be attacked.
Using artificial intelligence techniques to prevent vulnerabilities from being
introduced is a broad field and warrants an updated survey paper to identify
promising avenues of research in this area.

According to Cybersecurity Ventures, 111 billion lines of new software
code are created annually worldwide \cite{CodeSize}.  By utilizing automated mechanisms
to aid in detecting vulnerabilities before system deployment, a product team can focus 
more on feature development and performance. 
The large number of devices and applications being deployed today both increases the 
risks of vulnerabilities to a networked system and also provides a large collection of 
training data to use with artificial intelligence techniques. 
As we discuss papers in this survey, we will consider which areas of security 
vulnerability reduction are most likely to benefit from further
artificial intelligence advances in the near future.

The broadest recent survey in this area reviews papers from 2008-2015 and
explored how machine learning techniques were being applied to broad security
domains \cite{Jiang2016SoKAM}. Our survey differs from that work in 3 primary
ways. First, we will be including papers published as recently as 2018; second,
we will be considering all artificial intelligence approaches, not limiting
to machine learning; and third we will be focussed on vulnerability detection
and prevention, not attack detection or other security issues. 
Another recent survey paper is a student paper available on arXiv and discusses the 
DARPA sponsored Cyber Grand Challenge to detect, exploit, and/or repair vulnerabilities 
in software \cite{2017arXiv170206162B}.  We will discuss the results of these
surveys in more detail in sections ~\ref{sec:detect} and ~\ref{sec:repair}. In
our survey we include some references that are not peer-reviewed (such as
the student survey but also publications from companies). These references
are provided in the spirit of giving a broad understanding of the current
work in this area.

\begin{figure*}[h!tb]
\centering
\includegraphics[width=1.0\textwidth]{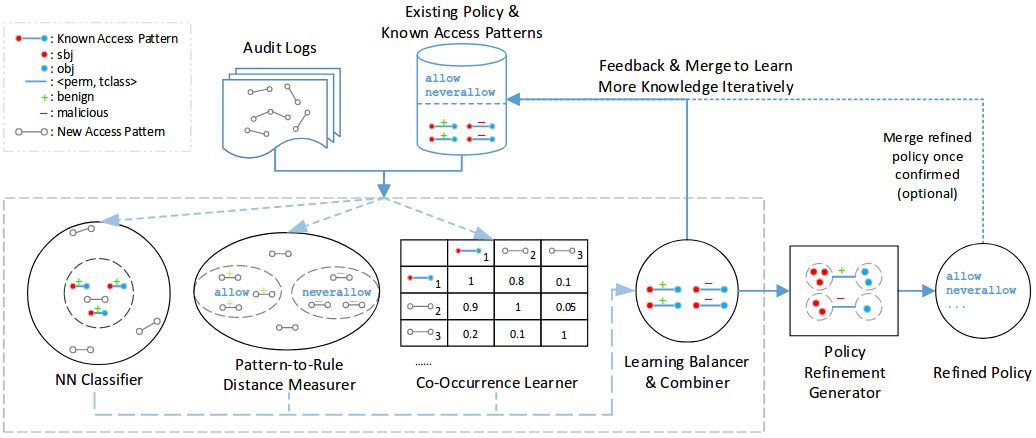}
\caption{Example learning system showing training inputs, learning components, and generated 
output. In this case, audit logs, access patterns, and existing access policies are used as 
inputs to generate refined policies. \cite{Jiang97}}
\label{fig:learning}
\end{figure*}

There are many ways to organize an algorithm to use AI with security, but figure
~\ref{fig:learning}, taken from one of our surveyed papers \cite{Jiang97}, is 
useful for understanding many common components. The figure identifies the input
data as audit logs, existing security access policies, and known access patterns. The learning
algorithm in the figure uses a nearest-neighbor classifier along with other
components to learn when policies should be loosened or tightened based on
the audit log information. The output of the system refines the policies. This
format of using training data to develop an AI system which can then be applied
to new inputs is a general approach used by many of the papers cited.

This paper is organized into 3 sections. Section ~\ref{sec:detect} covers
techniques to detect vulnerabilities in source code or binary code. Section ~\ref{sec:repair}
covers techniques to automatically repair vulnerabilities in source code.
Section ~\ref{sec:spec} discusses techniques to analyze security specifications
to insure the future code is correct or to check that access permissions match the documentation.
For each of the 3 areas, we will assess what we consider the most promising avenues of research.

\section{Vulnerability detection}
\label{sec:detect}

Vulnerability detection is the problem of identifying vulnerable code sequences
by analyzing source code prior to deployment. Table ~\ref{tab:detect} shows papers
which use AI techniques for vulnerability detection. Two of the papers focus on Android
apps and this highlights a primary reason for machine learning in the security space \cite{Jiang100}
\cite{Jiang95}. Companies whose reputation depends on their software quality are also
making platforms available to which large numbers of coders can contribute. The ability to
automatically process submitted applications greatly reduces the overhead of quality
assurance by trained professionals.

Yamaguchi, one of the authors cited in the table, wrote further about his techniques
in his PhD thesis \cite{Yamaguchi2015}. His work combines a new parsing strategy with
a joint data structure of program syntax, control flow, and data flow into a code
property graph. This graph is embedded into a vector space for use with machine learning
algorithms and mined for instances of vulnerable program patterns. For his application, he
found that supervised training for security was not as valuable as unsupervised training
based on clustering.  Yamaguchi was able to reduce the code needing human review by 95\% 
by focussing on dimensionality reduction, anomoly detection, and clustering.

\begin{table*}
\begin{tabular}{ |p{2cm}||p{2.5cm}|p{3.5cm}|p{6cm}| }
 \hline
 \multicolumn{4}{|c|}{Vulnerability detection} \\
 \hline
 Reference & Goal & AI Techniques & Comments \\
 \hline
\cite{Jiang94} & Limits of automated fingerprinting for OS & Decision tree, rule learner, SVM-SMO, instance-based clustering & ML tended to overfit on unimportant differences in variants. Training data was fuzz testing with candidate queries feeding into a classifier. \\
 \hline
\cite{Jiang100} & Score  and  rank  risks of Android apps & Naive Bayes, Probabilistic generative model & Analyzes GooglePlay apps and a Malware dataset based on permissions requested (READ\_SMS, ACCESS\_WIFI\_STATE, etc) to create single risk score to simplify user interactions. \\
\hline
\cite{Jiang95} & Identify sources and sinks from code of any Android API & Linear SVM & Recognizes sensitive data sources and potentially untrustworthy data sinks which might leak data by analyzing source code. \\
\hline
\cite{Mokhov14} & Identify bad coding practices & NLP, Modular A* Recognition Framework & Analyze CVE code base for training examples of weak code and use model to identify risk areas in source code.  \\
\hline
\cite{Jiang101} & Infer vulnerability search  patterns  in  C code & Complete-linkage clustering & Detect attacker-controlled data attack to a sensitive data sink. (Heartbleed is an example). Constructs code property graph and uses ML to reduce amount of code needed to audit by 95\%. \\
\hline
\cite{Jiang102} & Predict   future   data leak   instances   from network logs & Random Forest & Predict future organization breaches by analyzing externally measurable network features (misconfigured DNS, phishing sourced from organization, etc). \\
\hline
\cite{not-all-bytes-are-equal} & Improve fuzzing of inputs & Neural network & Fuzzing is a technique to provide crafted malicious input to test a program for vulnerabilities. This paper uses neural networks to learn which areas in programs to attack to test for vulnerabilities. \\
\hline
\cite{Phish} & Discover vulnerabilities in binary code & Symbolic execution and fuzzer & Searches for tainted input vulnerabilities. Avoids path explosion of symbolic execution by intelligently combining it with fuzzing techniques. \\
\hline
\end{tabular}
\caption{Artificial Intelligence techniques addressing automatic detection of software vulnerabilities}
\label{tab:detect}
\end{table*}

Two of our references in this category deal with Android applications \cite{Jiang100} \cite{Jiang95}. 
Given the large number of applications submitted for use on Android platforms, automating any
part of the security problem is valuable. One of these papers includes data showing that malicious
software frequently requests many more permissions than valid software (for example, only 2-3\%
of valid applications request the READ\_SMS permission, but almost 60\% of malware requests it) \cite{Jiang100}.
By training a system to recognize the malware request patterns, the 2-3\% of valid applications
are mostly classified as safe, while the malware is detected. The other Android reference compares
their technique for detecting unsafe data leaks to other dynamic taint analysis tools like
Fortify SCA and TaintDroid \cite{Jiang95}. In taint tracking, one monitors the flow of
data between resources such as the file system or network. The approach uses supervised training
to learn when privacy information such as location information or address book data is
leaked to risky destinations such as certain web domains. Their code was able to find
new data destinations used by malware (sinks) in the Android application set which were
not previously on lists of sinks for the ecosystem.

Microsoft has an enormous code base that must be maintained and developed
with attention to security vulnerabilities. Fuzzing is a dynamic
program technique which mutates program inputs to find vulnerabilities 
that cause crashes, buffer overflows, memory errors, and exceptions. Long Short
Term Memories (LSTMs) are used in recurrent neural networks to facilitate
learning patterns in time as well as syntax structures in human or computer languages.
Microsoft researchers used neural networks including LSTMs to learn patterns from 
past fuzzing explorations to guide future fuzzing mutations efficiently \cite{not-all-bytes-are-equal}.
Crashing a system is a key measurement of fuzzing efficacy and using their
LSTM guided fuzzing technique Rajpal et al., were able to find 37\% more unique crashes
than the benchmark fuzzing technique they compared against.

Automating the processes of checking code for vulnerabilities is important. DARPA 
sponsored the Cyber Grand Challenge to detect, exploit, and/or repair vulnerabilities in software;
and the Mayhem and Mechanical Phish \cite{Phish} programs were the top 2 ranked submissions. Both
used symbolic execution to aid in the search for vulnerabilities \cite{2017arXiv170206162B}.
Symbolic execution is a technique where input values are tracked symbolically through
a program, as opposed to running multiple test cases with specifically chosen values
for input parameters. Both Mayhem and Mechanical Phish are designed to discover
exploitable vulnerabilities in binary code. Their techniques can help uncover buffer
overflow vulnerabilities, format strings vulnerabilities, and general memory corruption
vulnerabilities.  However, neither winner is yet ready to be used for large, complex
systems and so further research in this area is warranted.

Using NLP techniques on code has shown interesting results \cite{Mokhov14}. Mokhov et al. show
how to teach a language model what bad code looks like using CVE and CWE cases and then use the
trained model to identify sections of source code that match given vulnerabilities. They build
their system on top of MARFCAT, a machine learning code analysis application they also
developed \cite{MARFCAT}. Their core NLP methodology includes 2 character n-grams with smoothing
techniques, which means the language is analyzed using these 2 character patterns but with
some adjustment made to smooth out the probabilities for rarely seen tuples. With appropriate
training their model can work on binary code or source code in various languages. MARFCAT learns 
known weaknesses by computing various language models from the CVE-selected test cases and 
then can compare these models to unseen code fragments for vulnerability discovery.

We conclude this section with reference to a cautionary paper. Remote operating system
fingerprinting in order to identify different OS versions usually requires manual effort.
Richardson et al. \cite{Jiang94} tested a previously published promising technique to eliminate manual
intervention. The new technique used an approach similar to fuzz testing. They found that
when tested on 329 different machine instances the results with the new technique were poor. 
They suspected the system overfitted to behavioral differences that were not OS-specific and concluded
that manual expertise would still be needed for OS fingerprint generation. In general,
many of the AI techniques that show initial promise need to be tested in larger
production environments before being broadly deployed.

\section{Vulnerability repair}
\label{sec:repair}

The next step beyond detecting vulnerabilities is to attempt automatic repairs with artificial
intelligence techniques. The goal of code repair is to learn from examples how to transform
vulnerable code into non-vulnerable code. As this field progresses, millions of lines of
legacy code could be scrubbed to improve security very broadly. Even early progress in this
area can be very valuable. If a new class of vulnerability is found (such as Spectre),
training examples of changing vulnerable code to protected code can often be generated
quickly (for Spectre, compiling Visual C++ with older vs latest MSVC can create training
cases \cite{Spectre}). Creating a automated system that can transform code with the vulnerability
to clean code can allow large software repositories to be repaired efficiently.

Klieber and Snavely \cite{KliSna16} use code transformations to repair common types of bugs. 
For example, access control is one of the top ten vulnerabilities in the OWASP list
\cite{OWASP} and falls under CWE-285; their approach uses a first order logic solver to
prove access controls are safe and will add appropriate conditions if it is not. They analyze
the code to determine the intended (implied) access controls that should be available to
the user and can repair functions to limit access to only this implied set. As an example
from the paper, consider a collaborative document viewing/editing system which allows documents
to be viewed by members of a team but only edited by the author given normal UI access
options. The approach taught how to compute the intended access control policy as well
as the policy available to an attacker without restriction to the normal UI interface. 
When an attacker's access exceeds the intended access, the codebase is repaired to
provide the proper limitation.

Another approach to program repair is to search for code changes which pass security test cases \cite{Le:2017}.  Le et al. use a domain-specific language for input along with test cases and sets up a
search space to find quality repairs to bugs in code. The code is analyzed to find likely lines
at risk for bugs, adjust those lines to use symbolic execution, and analyzes the symbolic execution
given the inputs and outputs provided for quality assurance. Bugs which are found can then
often be repaired to produce the correct output given the input.

One of our references is an example of a more specific yet widely applicable use of AI for 
improving mandatory access control (MAC) for Android \cite{Jiang97}. The system uses audit
logs from existing applications and semi-supervised learning to train a network to generate
improved MAC policies as well as improved audit logs. The improved audit logs can then
be used to improve the MAC policies further. Figure ~\ref{fig:learning} is from this paper 
and shows how the components of this system interact.

As an example of ongoing work in industry, we include an article by Draper in this
table \cite{Draper}. Heartbleed is a famous security bug from 2014 involving buffer
overflow is the OpenSSL cryptography library. Using a neural network trained on 170,000
C/C++ projects from GitHub, their DeepCode architecture found and fixed Heartbleed bugs
when run on previously unseen programs.

\begin{table*}
\begin{tabular}{ |p{2cm}||p{2.5cm}|p{3.5cm}|p{6cm}| }
 \hline
 \multicolumn{4}{|c|}{Vulnerability Repair} \\
 \hline
 Reference & Goal & AI Techniques & Comments \\
 \hline
\cite{Jiang97} & Semi-supervised learning for Android MAC & Nearest neighbors(metric: semantic connectivity between known and unknown subjects) & Analyzes access patterns from system usage to improve Mandatory Access Control policies. \\
\hline
\cite{KliSna16} & Repair Missing Function-Level Access Control & First order logic solver & Uses first order logic solver to prove access controls are safe and, if not, add the appropriate condition. This vulnerability type is one of top 10 on OWASP list \cite{OWASP}. \\
\hline
\cite{Le:2017} & Repair vulnerable code using programming by examples & Inductive Synthesis, Symbolic Execution & Sets up search space to find quality repairs to code and tests candidate fixes using test cases. \\
\hline
\cite{Draper} & Uses ML to eliminate found vulnerabilities & Neural network & This is not a scientific paper, but their approach is described more in \cite{MUSE}. Learn over large code corpus how to repair programs. \\
\hline
\end{tabular}
\caption{Artificial intelligence techniques aimed at automatically repairing vulnerabilities}
\label{tab:repair}
\end{table*}

\section{Specification analysis}
\label{sec:spec}

Advances in natural language processing have given researches the opportunity to automatically
process vulnerability descriptions or product specifications to assess security risks. In 2017,
CVE logged almost 300 vulnerabilities per week \cite{CVE}, which puts enormous burden on
system administrators to evaluate, prioritize, and patch critical vulnerabilities. Bozorgi et al. \cite{Jiang99}, use machine learning to digest the CVE vulnerability labels and descriptions and
produce a single score that summarizes both the exploitability of the vulnerability and its
severity. The goal is that the single score can be used by system administrators to prioritize
work on patches for new vulnerabilities. Their technique relies on an NLP technique known as
'bag-of-words'. For example they record whether particular tokens like "buffer", "heap", or "DNS"
appear in specific text fields like "title", "solution", or "product name". Their classifier is about
90\% accurate, much more so than the 'Exploitability' score provided by CVSS.

To help address risk assessment in new mobile applications, WHYPER has been trained to recognize
sentences in a smartphone application description to infer the permissions an application should
need \cite{Jiang96}. For example, if the application description includes "You can share the yoga exercise to your
friends via Email and SMS", then this indirectly implies the application should have the
READ\_CONTACTS permission on the smartphone. Applications which ask for permissions not
implied in the description to the customer could be classified as suspicious.

\begin{figure}[h!tb]
\centering
\includegraphics[width=0.5\textwidth]{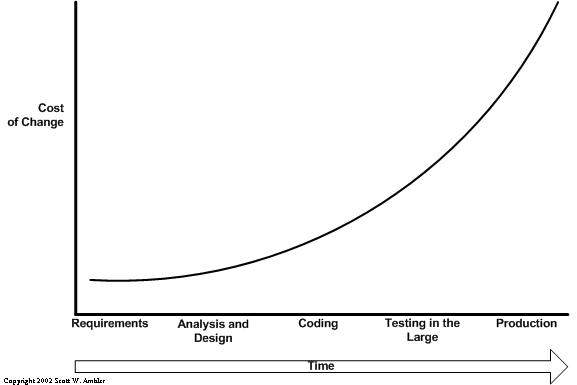}
\caption{Traditional cost of change curve. Agile coding arguably flattens this curve, but fixing bugs at requirements stage is clearly the least effort. \cite{Cost}}
\label{fig:cost}
\end{figure}

Our final paper in this category addresses the goal of finding vulnerability issues
in product specifications before code is being written. Figure ~\ref{fig:cost} shows the
traditional cost of change curve during software development.  As discussed in the introduction,
many artificial intelligence techniques applied to computer security are being applied
at the production stage - detecting attacks to running system by monitoring traffic or logs.
This paper aimed to highlight artificial intelligence techniques that can be applied
earlier in the design cycle. In order to find security vulnerabilities before code is even 
written, some researchers have looked at analyzing security requirements 
documents \cite{Malhotra16} \cite{Hayrapetian:2018}.
The earlier paper \cite{Malhotra16} is structured as a proposal for an approach to the problem.
That paper notes that previous work to detect requirement document problems focussed on 'weak 
phrases' and 'linguistic defects'. The paper proposes training a neural network on ISO security
standards and using the General Architecture for Text Engineering (GATE) to detect problems.
The paper proposes creating concept 
graphs and comparing the ISO standards to the document in question. For example, if the standards 
expect authentication to require a user ID and a password and the document only refers to a user 
ID, then the graph comparison would detect the missing password requirement and detect the error. 
The later paper is by some of the same authors as the first paper \cite{Hayrapetian:2018} and this 
paper provides results of the approach comparing security documents to ISO and OWASP standards.
They were able to reach 89\% accuracy in determining the completeness of standard statements, 
indicating a technique which could be a valuable tool for a specification reviewer, but
not yet a replacement for careful review.

\begin{table*}
\begin{tabular}{ |p{2cm}||p{2.5cm}|p{3.5cm}|p{6cm}| }
 \hline
 \multicolumn{4}{|c|}{Specification analysis} \\
 \hline
 Reference & Goal & AI Techniques & Comments \\
 \hline
\cite{Jiang99} & Predict possibility and timing of vulnerable exploits & Bag-of-words, Linear SVM & Assess impact and exploitability of new CVE and OSV vulnerabilities using bag-of-words and numerical ML techniques from vulnerability descriptions. \\
\hline
\cite{Jiang96} & Identify permission sentences in mobile apps & Part-of-speech(POS) tagging, phrase and clause parsing, named entity recognition, semantic graph, typed dependency & Uses NLP on application descriptions and determines if permissions given to app are justified. Bridges gap between user expectations and app functionality to reduce risk assessment of apps. \\
\hline
\cite{Hayrapetian:2018} & Evaluating security features in software requirements & NLP, Neural Network & Continuation of \cite{Malhotra16} work learning concept graphs from OWASP, OSI, and PCI specs and checks software documents for alignment to specs. \\
\hline
\end{tabular}
\caption{Artificial intelligence techniques aimed at discovering vulnerability risks in specifications}
\label{tab:spec}
\end{table*}

\section{Future work}
\label{sec:future}

As computer hardware continues to improve and more powerful artificial intelligence techniques 
are created, more such techniques will be applied to the problem of software security. 
To stay abreast of new developments, we recommend watching for future publications
by the authors and companies sited in this paper. Several of the frameworks proposed yield
promising but not yet productizable results, so further work in the areas discussed
is likely to be fruitful.

\section{Conclusion}
\label{sec:conclusion}

While the field of applying artificial intelligence to security vulnerabilities is
rather narrow, there is a lot of interest as shown by the diversity of papers
we have discussed in this survey. Some of the most intriguing results surveyed
include the ability to detect vulnerabilities in code and specification for new
smartphone applications, a widely visible area of risk to the consumer community.
The potential to automatically repair vulnerable code by learning the appropriate
code transformation could be very useful when enormous amounts of code need
to be reviewed due to the discovery of a new vulnerability. Also promising is the 
ability to analyze an english language specification for security risks with reference 
to security standards documents. Advances in natural language processing and 
code transformation techniques from fields outside security are creating 
approaches which can be applied to security problems. Given the estimated 111 billion lines
of software being generated annually in the world today, automated security techniques which
can be applied before systems are deployed will be critical for technology to
continue to be safely used by consumers, corporations, and governments.

\balance
{\small
\bibliographystyle{icml2018}
\bibliography{VulnAI}

\begin{thebibliography}{26}
\providecommand{\natexlab}[1]{#1}
\providecommand{\url}[1]{\texttt{#1}}
\expandafter\ifx\csname urlstyle\endcsname\relax
  \providecommand{\doi}[1]{doi: #1}\else
  \providecommand{\doi}{doi: \begingroup \urlstyle{rm}\Url}\fi

\bibitem[Agile()]{Cost}
Agile.
\newblock Examining the agile cost of change curve.
\newblock \emph{[Agile Modeling]}.
\newblock URL \url{http://www.agilemodeling.com/essays/costOfChange.htm}.

\bibitem[Blum et~al.(2017)Blum, Rajpal, and Singh]{not-all-bytes-are-equal}
Blum, William, Rajpal, Mohit, and Singh, Rishabh.
\newblock Not all bytes are equal: Neural byte sieve for fuzzing.
\newblock \emph{arXiv}, November 2017.
\newblock URL
  \url{https://www.microsoft.com/en-us/research/publication/not-all-bytes-are-equal-neural-byte-sieve-for-fuzzing/}.

\bibitem[Bozorgi et~al.(2010)Bozorgi, Saul, Savage, and Voelker]{Jiang99}
Bozorgi, M., Saul, L.~K., Savage, S., and Voelker, G.~M.
\newblock Beyond heuristics: learning to classify vulnerabilities and predict
  exploits.
\newblock In \emph{ACM Knoweldge Discovery and Data Mining}, 2010.
\newblock URL \url{https://dl.acm.org/citation.cfm?id=1835821}.

\bibitem[{Brooks}(2017)]{2017arXiv170206162B}
{Brooks}, T.~N.
\newblock {Survey of Automated Vulnerability Detection and Exploit Generation
  Techniques in Cyber Reasoning Systems}.
\newblock \emph{ArXiv e-prints}, February 2017.
\newblock URL \url{https://arxiv.org/pdf/1702.06162.pdf}.

\bibitem[CVE(2018)]{CVE}
CVE.
\newblock Cve details - vulnerabilities by type.
\newblock \emph{[Online]}, 2018.
\newblock URL \url{https://www.cvedetails.com/vulnerabilities-by-types.php}.

\bibitem[Draper(2017)]{Draper}
Draper.
\newblock Draper gets computers to fix their own bugs.
\newblock \emph{[Online]}, 2017.
\newblock URL
  \url{http://www.draper.com/news/draper-gets-computers-fix-their-own-bugs}.

\bibitem[Hayrapetian \& Raje(2018)Hayrapetian and Raje]{Hayrapetian:2018}
Hayrapetian, Allenoush and Raje, Rajeev.
\newblock Empirically analyzing and evaluating security features in software
  requirements.
\newblock In \emph{Proceedings of the 11th Innovations in Software Engineering
  Conference}, ISEC '18, pp.\  9:1--9:11, New York, NY, USA, 2018. ACM.
\newblock ISBN 978-1-4503-6398-3.
\newblock \doi{10.1145/3172871.3172879}.
\newblock URL \url{https://dl.acm.org/citation.cfm?id=3172879}.

\bibitem[Jagannathan(2017)]{MUSE}
Jagannathan, Suresh.
\newblock Mining and understanding software enclaves (muse).
\newblock \emph{[Online]}, 2017.
\newblock URL
  \url{http://materials.dagstuhl.de/files/15/15472/15472.SureshJagannathan1.Slides.pdf}.

\bibitem[Jiang et~al.(2016)Jiang, Nagra, and Ahammad]{Jiang2016SoKAM}
Jiang, Heju, Nagra, Jasvir, and Ahammad, Parvez.
\newblock Sok: Applying machine learning in security - a survey.
\newblock \emph{CoRR}, abs/1611.03186, 2016.
\newblock URL \url{https://arxiv.org/pdf/1611.03186.pdf}.

\bibitem[Klieber \& Snavely(2016)Klieber and Snavely]{KliSna16}
Klieber, W. and Snavely, W.
\newblock Automated code repair based on inferred specifications.
\newblock In \emph{2016 IEEE Cybersecurity Development (SecDev)}, pp.\
  130--137, Nov 2016.
\newblock \doi{10.1109/SecDev.2016.037}.
\newblock URL
  \url{https://resources.sei.cmu.edu/asset_files/conferencepaper/2016_021_001_483599.pdf}.

\bibitem[Le et~al.(2017)Le, Chu, Lo, Le~Goues, and Visser]{Le:2017}
Le, Xuan-Bach~D., Chu, Duc-Hiep, Lo, David, Le~Goues, Claire, and Visser,
  Willem.
\newblock S3: Syntax- and semantic-guided repair synthesis via programming by
  examples.
\newblock In \emph{Proceedings of the 2017 11th Joint Meeting on Foundations of
  Software Engineering}, ESEC/FSE 2017, pp.\  593--604, New York, NY, USA,
  2017. ACM.
\newblock ISBN 978-1-4503-5105-8.
\newblock \doi{10.1145/3106237.3106309}.
\newblock URL \url{http://www.cs.cmu.edu/~clegoues/docs/legoues-esecfse17.pdf}.

\bibitem[Liu et~al.(2015)Liu, Sarabi, Zhang, Naghizadeh, Karir, Bailey, and
  Liu]{Jiang102}
Liu, Y., Sarabi, A., Zhang, J., Naghizadeh, P., Karir, M., Bailey, M., and Liu,
  M.
\newblock Cloudy with a chance of breach: Forecasting cyber security incidents.
\newblock In \emph{USENIX Security}, 2015.
\newblock URL
  \url{https://www.usenix.org/system/files/conference/usenixsecurity15/sec15-paper-liu.pdf}.

\bibitem[Malhotra et~al.(2016)Malhotra, Chug, Hayrapetian, and
  Raje]{Malhotra16}
Malhotra, R., Chug, A., Hayrapetian, A., and Raje, R.
\newblock Analyzing and evaluating security features in software requirements.
\newblock In \emph{2016 International Conference on Innovation and Challenges
  in Cyber Security (ICICCS-INBUSH)}, pp.\  26--30, Feb 2016.
\newblock \doi{10.1109/ICICCS.2016.7542334}.
\newblock URL \url{http://ieeexplore.ieee.org/document/7542334/}.

\bibitem[Microsoft(2018)]{Spectre}
Microsoft.
\newblock Spectre mitigations in msvc.
\newblock \emph{[Visual C++ team blog]}, 2018.
\newblock URL
  \url{https://blogs.msdn.microsoft.com/vcblog/2018/01/15/spectre-mitigations-in-msvc/}.

\bibitem[Mokhov et~al.(2015)Mokhov, Paquet, and Debbabi]{MARFCAT}
Mokhov, S.~A., Paquet, J., and Debbabi, M.
\newblock Marfcat: Fast code analysis for defects and vulnerabilities.
\newblock In \emph{2015 IEEE 1st International Workshop on Software Analytics
  (SWAN)}, pp.\  35--38, March 2015.
\newblock \doi{10.1109/SWAN.2015.7070488}.

\bibitem[Mokhov et~al.(2014)Mokhov, Paquet, and Debbabi]{Mokhov14}
Mokhov, Serguei~A., Paquet, Joey, and Debbabi, Mourad.
\newblock The use of nlp techniques in static code analysis to detect
  weaknesses and vulnerabilities.
\newblock In Sokolova, Marina and van Beek, Peter (eds.), \emph{Advances in
  Artificial Intelligence}, pp.\  326--332, Cham, 2014. Springer International
  Publishing.
\newblock ISBN 978-3-319-06483-3.
\newblock URL
  \url{https://link.springer.com/chapter/10.1007/978-3-319-06483-3_33}.

\bibitem[OWASP(2017)]{OWASP}
OWASP.
\newblock Owasp top 10 - 2017.
\newblock \emph{[Online]}, 2017.
\newblock URL
  \url{https://www.owasp.org/images/7/72/OWASP_Top_10-2017_%28en%29.pdf.pdf}.

\bibitem[Pandita et~al.(2013)Pandita, Xiao, Yang, Enck, and Xie]{Jiang96}
Pandita, R., Xiao, X., Yang, W., Enck, W., and Xie, T.
\newblock Whyper: Towards automating risk assessment of mobile applications.
\newblock In \emph{USENIX Security}, 2013.
\newblock URL
  \url{https://www.usenix.org/system/files/conference/usenixsecurity13/sec13-paper_pandita.pdf}.

\bibitem[Peng et~al.(2012)Peng, Gates, Sarma, Li, Qi, Potharaju, Nita-Rotaru,
  and Molloy]{Jiang100}
Peng, H., Gates, C., Sarma, B., Li, N., Qi, Y., Potharaju, R., Nita-Rotaru, C.,
  and Molloy, I.
\newblock Using probabilistic generative models for ranking risks of android
  apps.
\newblock In \emph{ACM Conference on Computer and Communications Security},
  2012.
\newblock URL \url{https://nds2.ccs.neu.edu/papers/android_risks.pdf}.

\bibitem[Rasthofer et~al.(2014)Rasthofer, Arzt, and Bodden]{Jiang95}
Rasthofer, S., Arzt, S., and Bodden, E.
\newblock A machine-learning approach for classifying and categorizing android
  sources and sinks.
\newblock In \emph{Network and Distributed System Security}, 2014.
\newblock URL \url{http://www.bodden.de/pubs/rab14classifying.pdf}.

\bibitem[Richardson et~al.(2010)Richardson, Gribble, and Kohno]{Jiang94}
Richardson, D.~W., Gribble, S.~D., and Kohno, T.
\newblock The limits of automatic os fingerprint generation.
\newblock In \emph{ACM Conference on Computer and Communications Security},
  2010.
\newblock URL \url{https://www.gribble.org/papers/aisec05-richardson.pdf}.

\bibitem[Shoshitaishvili et~al.(2018)Shoshitaishvili, Bianchi, Borgolte, Cama,
  Corbetta, Disperati, Dutcher, Grosen, Grosen, Machiry, Salls, Stephens, Wang,
  and Vigna]{Phish}
Shoshitaishvili, Y., Bianchi, A., Borgolte, K., Cama, A., Corbetta, J.,
  Disperati, F., Dutcher, A., Grosen, J., Grosen, P., Machiry, A., Salls, C.,
  Stephens, N., Wang, R., and Vigna, G.
\newblock Mechanical phish: Resilient autonomous hacking.
\newblock \emph{IEEE Security and Privacy}, 16\penalty0 (2):\penalty0 12--22,
  March/April 2018.
\newblock ISSN 1540-7993.
\newblock \doi{10.1109/MSP.2018.1870858}.
\newblock URL \url{doi.ieeecomputersociety.org/10.1109/MSP.2018.1870858}.

\bibitem[Ventures(2017)]{CodeSize}
Ventures, Cybersecurity.
\newblock Application security report 2017.
\newblock \emph{[Cybersecurity Ventures]}, 2017.
\newblock URL
  \url{https://cybersecurityventures.com/application-security-report-2017/}.

\bibitem[Wang et~al.(2015)Wang, Enck, Reeves, Zhang, Ning, Xu, Zhou, and
  Azab]{Jiang97}
Wang, R., Enck, W., Reeves, D., Zhang, X., Ning, P., Xu, D., Zhou, W., and
  Azab, A.~M.
\newblock Easeandroid: Automatic policy analysis and refinement for security
  enhanced android via large-scale semi-supervised learning.
\newblock In \emph{USENIX Security}, 2015.
\newblock URL
  \url{https://www.usenix.org/system/files/conference/usenixsecurity15/sec15-paper-wang-ruowen.pdf}.

\bibitem[Yamaguchi et~al.(2015)Yamaguchi, Maier, Gascon, and Rieck]{Jiang101}
Yamaguchi, F., Maier, A., Gascon, H., and Rieck, K.
\newblock Automatic inference of search patterns for taint-style
  vulnerabilities.
\newblock In \emph{IEEE Symposium on Security and Privacy}, 2015.
\newblock URL
  \url{https://user.informatik.uni-goettingen.de/~krieck/docs/2015-ieeesp.pdf}.

\bibitem[Yamaguchi(2015)]{Yamaguchi2015}
Yamaguchi, Fabian.
\newblock \emph{Pattern-Based Vulnerability Discovery}.
\newblock PhD thesis, Georg-August University School of Science, 2015.
\newblock URL
  \url{https://ediss.uni-goettingen.de/bitstream/handle/11858/00-1735-0000-0023-9682-0/mainFastWeb.pdf}.

\end{thebibliography}
}


\end{document}